\documentclass[aps,pre,fleqn,showpacs,superscriptaddress,twocolumn,reprint]{revtex4-2}
\usepackage{textcomp}
\usepackage{bm}
\usepackage{amsmath,amsfonts,amssymb}
\usepackage{color}
\usepackage{hyperref}
\usepackage{listings}
\usepackage{float}
\usepackage{graphicx}
\usepackage{enumitem}
\usepackage[capitalise]{cleveref}
\usepackage{overpic}
\usepackage{CJKutf8}
\usepackage{verbatim}
\usepackage{epstopdf}
\usepackage{ulem}
\usepackage{makecell}

\emergencystretch=\maxdimen
\hypersetup{
  colorlinks,
  backref,
  pdfborder=false
}

\newcommand{\CUB}{Renewable and Sustainable Energy Institute, University of Colorado, Boulder, Colorado 80309, USA}
\newcommand{\LRC}{Lodestar Research Corporation, 360 Bobolink Court, Louisville, Colorado 80027, USA}
\newcommand{\PPPL}{Princeton Plasma Physics Laboratory, Princeton, New Jersey 08543, USA}

\begin{document}
\title{Transport Barrier and Spinning Blob Dynamics in the Tokamak Edge}

\author{Junyi \surname{Cheng}}
\affiliation{\CUB}

\author{James \surname{Myra}}
\affiliation{\LRC}

\author{Seung-Hoe \surname{Ku}}
\affiliation{\PPPL}

\author{Robert \surname{Hager}}
\affiliation{\PPPL}

\author{Choong-Seock \surname{Chang}}
\affiliation{\PPPL}

\author{Scott \surname{Parker}}
\affiliation{\CUB}

\date{\today}

\begin{abstract}

In this work, we investigate the dynamics of plasma blobs in the edge of magnetic confinement devices using a full-f \mbox{gyrokinetic} particle-in-cell code with X-point geometry. In simulations, the evolution of a seeded blob is followed as it approaches a naturally-forming zonal shear layer near the separatrix, where the blob is stabilized by a large spin induced by the self-consistent adiabatic electron response, and blob bifurcation and trapping are observed during the cross-field propagation of blobs. A new theoretical explanation in both the zonal free and zonal shear layer is constructed, where the dominant $\mathbf{E}\times\mathbf{B}$ spin motion is included. A theoretical condition for a transport barrier induced by the interaction between spinning blobs and the zonal shear layer is obtained, and its scaling is verified with simulations. The new theoretical framework, especially the transport barrier, is applicable to explain and predict various experimental phenomena. In particular, the transport barrier condition calculated with experimental parameters demonstrates that the blob radial transport for H mode is smaller than L mode in experiments.
\end{abstract}

\maketitle
\sloppy

Coherent structures having poloidally and radially localized density structures and extended along the magnetic field, known as plasma blobs, often dominate the cross-field transport of particles and energy in the edge region between the confined plasma and material surface in magnetic confinement devices. The cross-field propagation of plasma blobs is of intrinsic interest for a better understanding of edge transport in future toroidal fusion devices and has drawn significant attention from the physics community \cite{KrasheninnikovPLA2001, KrasheninnikovJPP2008, D'IppolitoPoP2011, ChangNF2017, LiNF2020, PittsNME2019, CarraleroNF2017, VianelloNF2020, WynnNF2018, MilitelloPPCF2016}. Furthermore, we will see that rapidly spinning blobs, considered here, have vortex interactions with a sheared flow layer that are of fundamental interest, reminiscent of similar phenomena in 2D neutral fluids.

In the \mbox{conventional} paradigm\cite{KrasheninnikovPLA2001} of convective blob transport, the charge separation induced by opposite ion and electron vertical drift leads to a dipole potential and such a potential leads to an outward $\mathbf{E}\times \mathbf{B}$ drift. Although much progress  has been made in understanding the nonlinear dynamics of these structures \cite{KrasheninnikovJPP2008,D'IppolitoPoP2011}, realistic experimental conditions are complicated and further investigation is warranted. For example, the mushroom shape of blobs \cite{KatzPRL2008} due to charge separation has been demonstrated in a linear plasma device; however, blobs in the tokamak have a more stable nearly circular cross-section\cite{ZwebenPPCF2016,ZwebenPoP2022} shape. Blobs with negative cross-field velocity (radial velocity in a tokamak) have been observed \cite{ZwebenPPCF2016,TsuiPoP2018}. The blob emission in the SOL (scrape-off layer) region in H mode is less than L mode, which is observed in different tokamaks, such as NSTX\cite{BoedoPoP2001} and DIII-D\cite{ZwebenPoP2022}.

In this work, we investigate the blob dynamics by carrying out simulations using the edge-physics-aimed 5D particle-in-cell (PIC) gyrokinetic code XGC \cite{ChangPRL2017, KuPoP2018}. The blobs are seeded inside the separatrix using a realistic magnetic equilibrium from a C-Mod H-mode discharge\cite{HughesNF2018}. We consider the interaction of blobs born just inside the separatrix with a naturally forming shear layer near the separatrix. In the simulations, blobs with a large amplitude are stable with large spin and small cross-field motion. It is found that under certain conditions the spinning blob bifurcates, or splits, into two distinct parts at the midplane where the inner part is trapped in the core region, and the outer part moves outward radially into the SOL region. Hence, the $E_r$ shear layer near the separatrix acts as a transport barrier.
A new theory that includes the dominant $\mathbf{E}\times\mathbf{B}$ spin motion in the parallel Ohm’s law, vorticity and continuity equations is required. The potential coming from the adiabatic electron response giving rise to the spin is treated as zero-order, whereas the dipole potential is first-order. The blob can be stabilized by spinning and distorted by the large $E_r$ shear near the separatrix. The blob's radial motion responds to the zonal electrostatic potential well with the blob being charged positive, consistent with the simulation results. The new theory may explain the negative radial velocity of blobs seen in experiments. The blob bifurcation in the separatrix region is also observed in experiments, and the transport barrier condition obtained here and calculated with experimental parameters demonstrates that the blob radial transport for H mode is smaller than L mode in experiments.
\\

\textbf{Simulation results.} Here, the electrostatic version \cite{HagerJCP2016,KuPoP2018} of XGC is used for simulations, which employs drift kinetic electrons and gyrokinetic ions with a logical sheath boundary condition \cite{ParkerJCP1993} at the divertor plate. For simplicity, the source term is turned off, and no neutral-plasma interaction is included. The geometry and magnetic equilibrium are based on a lower single null Alcator C-Mod EDA high (H) mode discharge \cite{HughesNF2018} $\#1160930033$ with a plasma current of $I_p=1.4 \mathrm{MA}$, a toroidal magnetic field on the axis of $5.8\mathrm{T}$, an input power of $5.4\mathrm{MW}$, and a record volume averaged pressure exceeding $0.2 \mathrm{MPa}$. We simplify the problem in two ways. First, we set the density and temperature profiles to be uniform to avoid complications of gradient-driven instabilities super-imposed on the seeded blob physics. Second, in order to reduce the required computational resources, we reduce the magnetic field by a factor of $8$, resulting in a magnetic field magnitude of $0.6 \mathrm{T}$ at the separatrix at the outer-most midplane. (Experimental conditions will be restored in our final scaling estimates.) The ion grad-$B$ drift is directed downwards. The ions and electrons have a uniform temperature profile $T_e=T_i=60\mathrm{eV}$ and the density profile is uniform $n_e=n_i=10^{19}\mathrm{m^{-3}}$. Note, since the gradients of blob density and the profile satisfy $\nabla \mathrm{ln} n_b>\nabla\mathrm{ln}n_0$, the turbulence generated by profile gradients is suppressed. Thus simulations here are used only to investigate the blob physics.

First, we carry out a simulation without blobs as a controlled case. No turbulence is observed. A small amplitude zonal potential develops at the separatrix and the flux surface at the ``virtual'' upper X-point due to the charge absorption at the divertor. This controlled simulation verifies turbulence is suppressed with uniform density and temperature profiles.

\begin{figure}[!ht]
\includegraphics[width=0.132\textwidth,trim=0 0 0 0 0, clip]{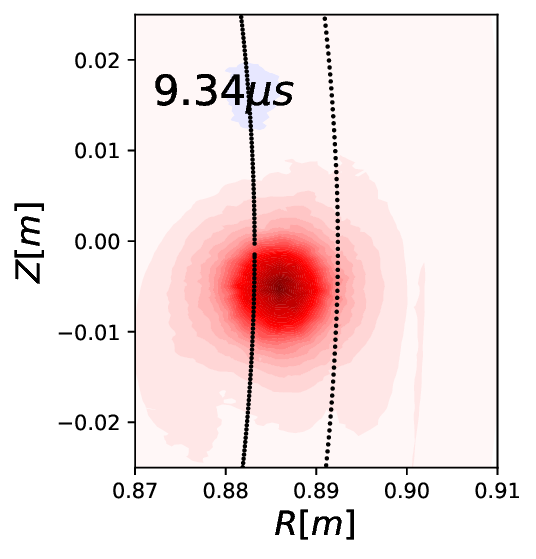}
\includegraphics[width=0.105\textwidth,trim=5 0 0 0 0, clip]{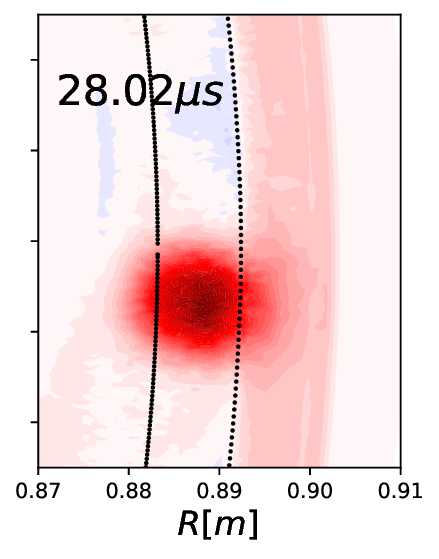}
\includegraphics[width=0.105\textwidth,trim=5 0 0 0 0, clip]{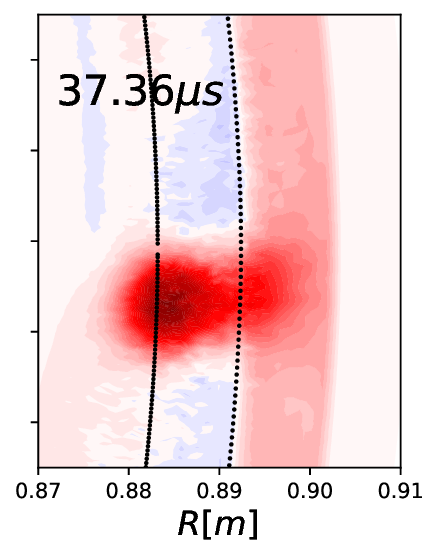}
\includegraphics[width=0.105\textwidth,trim=5 0 0 0 0, clip]{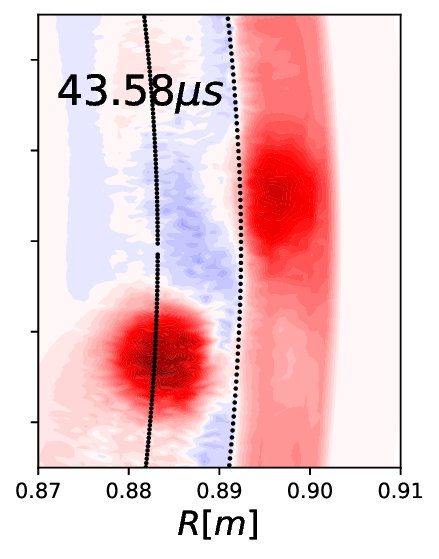}
\caption{The blob structure of the electrostatic potential $\phi$ in the $RZ$ plane at different times. 
 The left line indicates the flux surface where the blob is initialized and the right line indicates the separatrix surface. For instance, at $t=28.02\mathrm{\mu s}$, the blob radial velocity $v_b^R\approx 76.16\mathrm{m/s}$, the blob spin velocity $v_{s}\approx 11565.12\mathrm{m/s}$, the averaged blob perturbed density $\delta n$ is about $0.51$ times of background density, and the adiabaticity $e\phi/T/(\delta n/n)\lesssim 1$. }\label{fig:snap}
\end{figure}

\begin{figure}[!ht]
\includegraphics[width=0.35\textwidth]{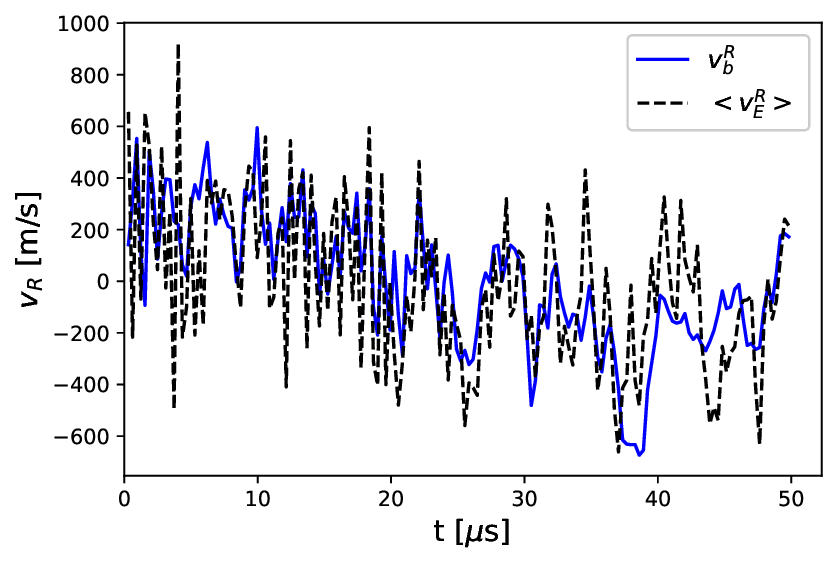}
\caption{The blob center-of-mass radial velocity $v_b^R$ versus time compared to the averaged radial component of
 $\mathbf{E}\times \mathbf{B}$ drift $\langle v_E^R \rangle$ versus time. Averaging $\langle ... \rangle$ is done over the blob electron density.}\label{fig:vb}
\end{figure}

We now carry out a simulation where a blob is seeded in the closed-flux region just inside the separatrix at $\psi/\psi_x=0.94$, where $\psi$ is the poloidal flux and $\psi_x$ is the poloidal flux at the separatrix.   Our simulations are not intended to model the main pedestal where Reynolds stress and plasma profile gradients play an important role. The radial size of the seeded blobs is $3.71$ times the ion gyro radius. The structure of the seeded blob is Gaussian in the radial coordinate and the other bi-normal coordinate and a cosine function along the field line. The amplitude of density perturbation is an input parameter. When the density perturbation is small, the blob has an obvious charge separation and a radial outward motion; when the density perturbation is large enough, as shown in \cref{fig:snap}, the blob structure is stable due to a fast spin induced by the adiabatic electron response. The blob moves radially outward in the closed-flux region at a speed much smaller than the spin velocity. This behavior is expected from previous fluid studies in the SOL in simplified geometry\cite{MyraPoP2004,AngusPoP2012}. There is also a small downward drift until the blob's radial motion stops. Then the blob bifurcates into two: one blob stays inside the separatrix and continues to drift downward; the other smaller blob moves across the separatrix. Hence, the separatrix region acts as a transport barrier for the inner blob.

The blob radial velocity $v_b^R$ can be measured by taking the time derivative of the center-of-mass\cite{WiesenbergerPoP2014} location at each time step. The radial component of the $\mathbf{E}\times \mathbf{B}$ drift averaged over the blob electron density, $\langle v_E^R \rangle$, indicates that the radial motion is caused by the average $\mathbf{E}\times \mathbf{B}$ drift. As shown in \cref{fig:vb}, $v_b^R$ and $\langle v_E^R \rangle$
are in reasonable agreement. The dipole potential structure in the vertical direction must still 
exist since it causes a pure radial $\mathbf{E}\times \mathbf{B}$ radial motion. However, it is not visible in \cref{fig:snap} because the adiabatic potential dominates. In what follows, the dipole potential structure will be treated as a first-order expansion of the zero-order positive potential due to the adiabatic electron response. 
\\

\textbf{Theoretical explanation of results.} We start with the usual fluid vorticity equation
\begin{equation}\label{vor0}
\frac{\omega_{pi}^2}{4\pi\Omega_i^2}\left(\frac{\partial}{\partial t}+\mathbf{v}_E\right)\nabla^2_\perp\phi=\nabla J_\|+\frac{2}{B}\mathbb{b}\times\kappa\cdot\nabla p,
\end{equation}
and the parallel Ohm's law  
\begin{equation}\label{ohm}
\nu J_\|+\mathbf{v}_E\cdot\nabla J_\|=\frac{ne^2}{m_e}\left(\frac{T_e}{e}\nabla_\|\mathrm{In}n-\nabla \phi\right),
\end{equation}
along with the fluid continuity equation, which is used to eliminate density and get $J_\parallel (\phi)$.
Note that the spin induced by $\mathbf{E}\times \mathbf{B}$ drift dominates the blob motion, thus the parallel Ohm's law \cref{ohm} and the fluid equations include the $\mathbf{E}\times \mathbf{B}$ and spin motion in leading order. Here $\omega_{pi}$ is the ion plasma frequency, $\Omega_i$ is the ion gyrofrequency, $\mathbf{v}_E=\mathbf{E}\times \mathbf{B}/B^2$ is the $\mathbf{E}\times \mathbf{B}$ drift, $\phi$ is the potential, $J_\parallel$ is the parallel current, $\kappa=\mathbf{b}\cdot\nabla\mathbf{b}$ is the magnetic curvature, $p$ is the pressure, $\nu$ is the collision frequency, $n$ is the total density and $\nabla_\parallel=\mathbf{b_0}\cdot\nabla$.

Since the adiabatic electron response dominates the zero-order potential and the dipole potential appears in the next order, 
we solve the vorticity equation for the azimuthally symmetric zero order $(\phi_0, m=0, k_\parallel=0)$ and first order dipole $(\phi_1, m=1)$ component of the potential in the local cylindrical blob coordinates where 
$
\phi \propto \sum e^{-im\theta+ik_\|z}.
$
Here, $\theta$ is the local azimuthal angle ($\theta = 0$ in the $e_R$ direction at the outboard midplane) relative to the blob center and $z$ indicates the coordinate along the field line.

In zero order, assuming adiabatic electrons, the right-hand side of Eq. (2) and hence the parallel current vanishes.  In the first order, the $\kappa$ term in Eq. (1) acts as a source term for the m = 1 dipole potential. After some algebra that combines the m = 1 first-order vorticity, density, and parallel current equations, one obtains 

\begin{equation}\label{vor1}
L_1\phi_1-k_\parallel^2\sigma_{\parallel,1}\phi_1=\frac{2c}{B}\mathbf{b}\times\kappa\cdot\nabla p,
\end{equation}
where $L_1$ is an inertial term given by
$
L_1=\frac{\omega_{pi}^2}{4\pi\Omega_i^2}\left(-i\Omega_b\nabla_\perp^2+\frac{c}{B}\nabla(\nabla^2_\perp\phi_0)\cdot\mathbf{b}\times\nabla\right),
$
and $\sigma_{\parallel,1}$ is the conductivity relating $J_{\parallel,1}$ to $E_{\parallel,1}$
\begin{equation}\label{simga1}
\sigma_{\parallel,1}=i\frac{ne^2}{m_e}\frac{1-\alpha_{ad}}{\Omega_b+i\nu-k_\parallel^2 v_{te}^2/\Omega_b},
\end{equation}
Here $\Omega_b=\frac{c}{rB}\frac{d\phi_0}{dr}$ is the spin frequency, $v_{te}$ is the thermal velocity and $\alpha_{ad}=\frac{T_e}{en_0}\frac{dn_0}{d\phi_0}$ indicates the adiabaticity. The $\alpha_{ad}$ parameter is introduced for the theoretical description; the simulations employ the full drift-kinetic electron response.

Based on the above equations, we can solve for the propagation of spinning blobs driven by the curvature mechanism which is well known in \cref{vor0,ohm} to charge-polarize the blob. In the present formalism, when $\sigma_{\parallel,1}$ is real (for example in the collisional limit) $\mathrm{Re}[\phi_1 \mathrm{exp}(-im\theta)] \propto \mathrm{cos}\theta$ results in an electric field in the $e_Z$ direction (up-down in the $RZ$ plane) and hence a radially outward $\mathbf{E} \times \mathbf{B}$ blob propagation velocity. Poloidal blob propagation results from \cref{vor1,simga1} when the left-hand side of \cref{vor1} gives $\phi_1$ an imaginary part.  In addition to dipole charge polarization, the blob motion is also affected by the background $\mathbf{E}\times \mathbf{B}$ flow, curvature, and grad-$B$ drifts.
\\

\textbf{Effect of the zonal field.} To include the effect of the zonal field and the resulting sheared flows, $\phi$ contains zero order terms describing the background zonal flow $\phi_{0f}$ as well as the blob spin $\phi_0$. The calculation is performed in the frame of the flow at the blob center so that $\partial/\partial t$ can be neglected. The resulting equation for the first order response $\phi_1+\phi_{1f}$ from the curvature and flow terms respectively is then

\begin{equation}\label{vorf1}
    \begin{split}
        &(L_1-\sigma_{\parallel,1}k_\parallel^2)(\phi_1+\phi_{1f}) = \\
        &\frac{2}{c}\mathbf{b}\times\mathbf{\kappa}\cdot p-\frac{\omega_{pi}^2}{4\pi\Omega_i^2}(\mathbf{v}_{E0}\cdot\nabla\nabla_\perp^2\phi_{0f}+\mathbf{v}_{E0f}\cdot\nabla\nabla^2_\perp\phi_0)  \\
        &\sim \frac{ne}{\omega_i\delta_b}\left[\frac{\delta n}{n}\frac{2c_s^2}{R}+\frac{e\phi_0}{T_e}\left(\frac{\rho_s^2}{L_f^2}+\frac{\rho_s^2}{\delta_b^2}\right)\frac{eE_{0f}}{m_i}\right]
    \end{split}
\end{equation}

The first term on the right-hand side of \cref{vorf1} is the source of the charge separation process, the second term is the source of the zonal field effect. Here local Cartesian coordinates at the outboard midplane are employed where $x$ is radial and $y$ is binormal, $\mathbf{v}_{E0f}=v_{E0f}\mathbf{e}_y=(c/B)\mathbf{b}\times\nabla\phi_{0f}$, and a small flow ordering $\nabla v_{E0f}\ll \Omega_b$ has been assumed to neglect quadratic products of flow terms and to simplify the operators on the left-hand side. Near the blob center at $x = 0$, $v_{E0f}\approx x\partial v_{E0f}/\partial x\sim \textrm{min}(\delta_b,L_f)v_{E0f}/L_f$, where $\delta_b$ is the blob radius, assumed to be circular, and $L_f$ is the scale length of the flow shear.

The size estimate of the terms on the right-hand side of  \cref{vorf1} expresses the sources in terms of the charge polarizing force per unit mass acting on the spinning blob.  As is well known\cite{KrasheninnikovJPP2008} the curvature and grad-$B$ effects give rise to the effective gravitational acceleration $g\sim 2c_s^2/R$. The factor $\delta n/n$ takes into account the blob amplitude over the background density\cite{D'IppolitoPoP2011}.

The flow term in square brackets has two parts. The second part can be expressed as $e\phi_0\rho_s^2eE_{0f}/(T_e\delta^2_bm_i)\sim(\Omega_b/\Omega_i)(eE_{0f}/m_i)$ which shows that the force on the blob is proportional to the blob vorticity $\bar{\omega_b}\equiv\Omega_b$ times the zonal electric field force $eE_{0f}/m_i$, i.e., the blob reacts to the zonal field like a (positive) charge since charge is proportional to vorticity through the Poisson equation. The actual interaction is complicated by the fact that in the flow frame, $v_{E0f}$ and $E_{0f}$ change signs across the blob center, so that the blob actually reacts to the velocity shear, as will be made explicit in \cref{gtb}.

The first term in square brackets can be rewritten as $(\bar{\omega}_f/\Omega_i)(e\phi_0/m_i\delta_b)(\delta_b/L_f)$ which shows the interaction of the blob electric field with the vorticity of the flow, $\bar{\omega}=v_{E0f}/L_f$. The extra factor of $\delta_b/L_f$ makes the two terms symmetric noting that one factor of $\delta_b$ has been pulled outside the square brackets.  

Both of these terms show explicitly the mutual interaction of two different sources of vorticity arising from the blob and the shear layer. Blob and shear layer interactions have been studied previously in slab geometry fluid simulations \cite{YuPoP2003,MyraPoP2004} and experiments\cite{SpolaorePoP2002,GrenfellNF2020}; however, the vortex analysis and interpretation turn out to be particularly transparent in the present large spin ordering.

The radial dependence of the zonal field, and the resulting shearing effect on blobs, compete with spin which can stabilize the blob and preserve its coherency. The stabilization criteria for spinning blobs from this competition is $\Omega_b>\nabla v_{E0f}\sim v_{E0f}/L_f$. The shearing rate can also affect the blob shape when it is large enough to distort the blobs as indicated in the simulations. For instance, at $37.36\mathrm{\mu s}$ in simulations, $\Omega_b\approx 1.70\times 10^6 \mathrm{s^{-1}}$ is less than $\nabla v_{E0f}\approx 1.95\times 10^6 \mathrm{s^{-1}}$, resulting in the blob bifurcation (\cref{fig:snap}). Thus, rapidly spinning blobs usually have a more stable shape. Blob stability is widely observed in experiments\cite{ZwebenPPCF2016,ZwebenPoP2022}, although direct experimental detection of blob spin has so far not been possible.
\\

\begin{figure}[!ht]\label{fig:potential}
\includegraphics[width=0.3\textwidth]{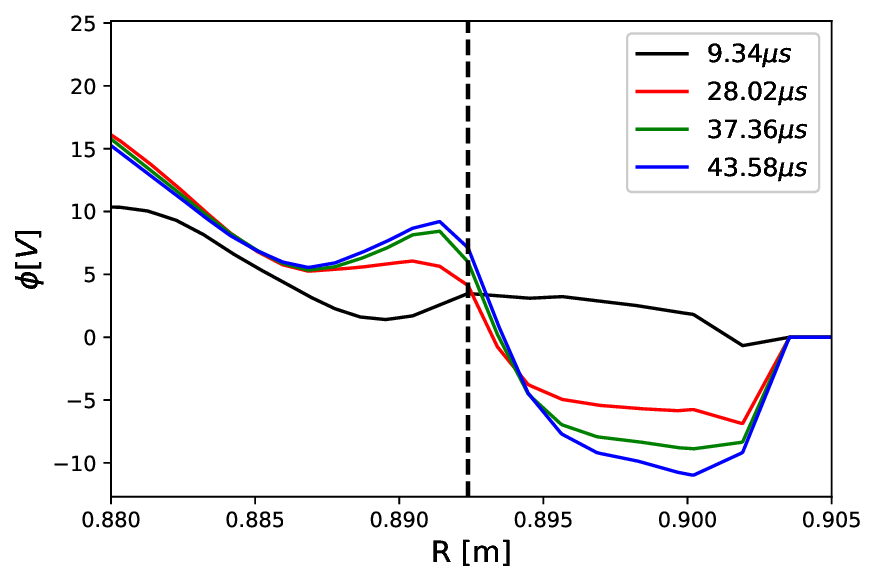}
\includegraphics[width=0.15\textwidth, trim=0 0 0 0 0, clip]{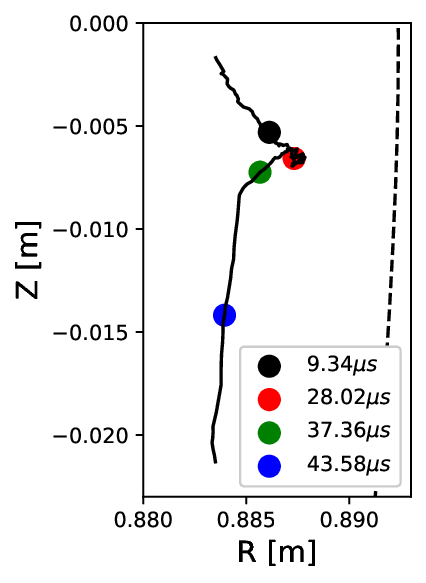}
\caption{The time evolution of zonal potential (left) and trajectories of the center-of-mass blob location (right). In the left figure, the solid lines are the zonal potential at different time and the black dashed line indicates the separatrix. In the right figure, the solid line illustrates the time history of the center-of-mass location in the $RZ$ plane; the colored dots indicate the center-of-mass locations at different times; the dashed line indicates the separatrix surface. Note that the zonal potential is due to particle loss in the divertor.}\label{fig:potential}
\end{figure}

\begin{figure}[!ht]\label{fig:scaling}
\includegraphics[width=0.32\textwidth]{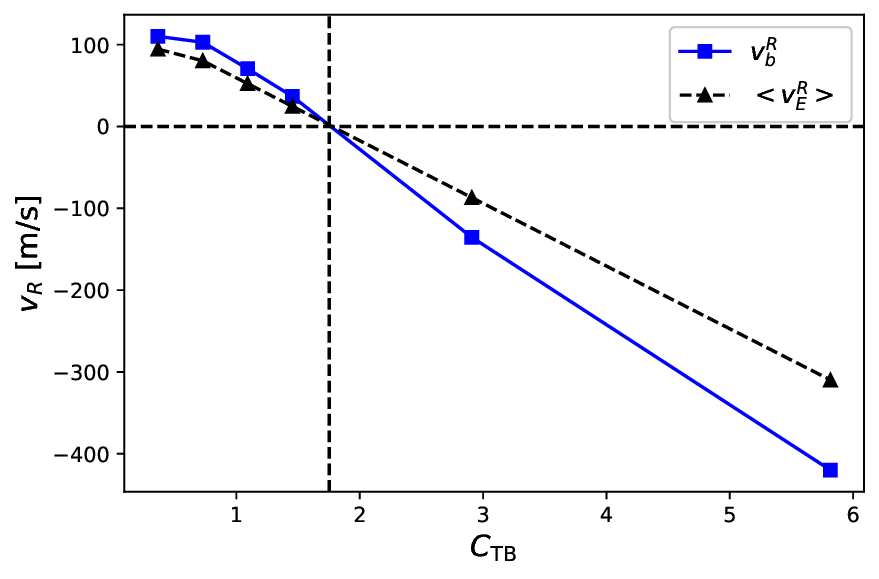}
\caption{Scaling of the transport barrier criteria $C_{\mathrm{TB}}$ versus radial velocity $v_b^R$ and averaged $\mathbf{\rm E}\times\mathbf{\rm B}$ velocity $<v_E^R>$. In the scaling, we restart the simulations at a selected time point $t=9.34\mu s$ with manually altered zonal fields $[0.5,1,1.5,2,4,8]\phi_{f}$, corresponding to different $C_{\mathrm{TB}}$. For $\phi_{f}$, $C_{\mathrm{TB}}\approx 0.727$ with $L_f\approx 19 \mathrm{mm}$, $\delta_b\approx 13\mathrm{mm}$, $v'_{E0f}\approx 4.40\times 10^5\mathrm{s^{-1}}$, $\Omega_i\approx 3.47\times 10^7 \mathrm{rad/s}$ and $R\approx 0.890\mathrm{m}$. The cross point of the vertical and horizontal dashed lines indicates the exact turning point $C_{\mathrm{TB}}\approx 1.753$ in the simulations.}\label{fig:scaling}
\end{figure}

\textbf{Blob transport barrier.} The radial drift velocity induced by dipole response due to the effect of the curvature drift charge separation is always outward, while the radial drift velocity induced by the zonal field effect depends on $E_{0f}$ and its shear, which can be negative or positive (\cref{fig:potential}). Thus, when these two mechanisms compete to determine the radial velocity, a transport barrier for the spinning blobs can be induced.
By comparing the flow terms on the right-hand side of \cref{vorf1} to the curvature term, it is possible to derive the condition for a blob transport barrier. Recall that when the left-hand side is real (dissipative) the curvature term drives the blob radially outward, across the separatrix and into the SOL.  The condition for the flow terms to dominate the curvature term is
\begin{equation}\label{gtb}
C_{\mathrm{TB}}=\left(\frac{1}{L_f}+\frac{1}{\delta_b}\right)\frac{R v'_{E0f}}{2\Omega_i}>1,
\end{equation}
where the estimate $v_{E0f}=\mathrm{min}(L_f,\delta_b)v'_{E0f}$ has been used to cancel one factor of $L_f$ or $\delta_b$ in the final expression. See the comment following in \cref{vorf1}. In deriving \cref{gtb} the condition $e\phi_0/T_e\sim \delta n/n$ for an adiabatic blob has been employed. As shown in \cref{fig:scaling}, for the simulation reported here, the inequality in \cref{gtb} is satisfied within factors of order unity, and the dynamics observed in the simulation agrees well with our theoretical expectation for net radial repulsion by the shear layer and shear-driven distortion and bifurcation of the blob. Furthermore, as discussed subsequently, the transport barrier condition is also relevant for experiments. 
\\

\textbf{Comparison with experiments.}
Comparison between theory, simulations, and experiments is a good tool to verify and validate the theory. We note the following qualitative and quantitative correspondences.
(a) The new theoretical framework does not conflict with the dipole potential structure observed in an experimental cross-correction analysis\cite{GrulkePoP2006}, since the dipole potential is also included in the theory.
(b) Blob bifurcation in the separatrix region has been clearly observed in C-Mod\cite{TerryPoP2003,TerryNF2005} and NSTX\cite{ZwebenNF2004}, which agrees with our simulations and theoretical expectation.
(c) Negative radial motion of blobs in TCV \cite{TsuiPoP2018} and NSTX \cite{ZwebenPPCF2016} is potentially due to their interaction with shear layers.
(d) The zonal field effect in the edge region is larger in H mode than in L mode. Thus blob emission into the SOL is expected and measured to be lower in H mode than in L mode, for example, in DIII-D \cite{BoedoPoP2001} and NSTX\cite{ZwebenPoP2022}. A 3kHz oscillation in the edge zonal field region of NSTX was found to be anti-correlated \cite{ZwebenPoP2010,SechrestPoP2011} with blob emission into the SOL. The above two experimental results are consistent with our theoretical prediction.
(e) According to the experimental data\cite{TerryPoP2003,McDermottPoP2009} in C-Mod, $L_f\approx 5\mathrm{mm}$, $\delta_b\approx 5\mathrm{mm}$, $v'_{E0f}\approx 2\times 10^6\mathrm{s^{-1}}$ for H mode, $5\times 10^5 \mathrm{s^{-1}}$ for L mode, $\Omega_i\approx 2.25\times 10^8 \mathrm{rad/s}$ and $R\approx0.89\textrm{m}$ in the separatrix region, $C_{\mathrm{TB}}\approx 1.09$ for H mode and $C_{\mathrm{TB}}\approx 0.27$ for L mode, which shows a good quantitative agreement.
(f) Recently a detailed study of blob-vortex interactions with shear layers in L and H mode has been carried out for the RFX-Mod device. The results\cite{GrenfellNF2020} are in qualitative agreement with the present theory.
Thus, the present theoretical framework for blob structure and motion is consistent with the experimentally observed behavior of blobs.
\\

\textbf{Summary.} In this letter, we investigated blob dynamics by seeding a blob as an initial condition in the comprehensive full-f XGC particle code with gyrokinetic ions, drift-kinetic electrons and realistic X-point geometry in the electrostatic limit. We find that spin dominates the blob motion due to the adiabatic electron response. The dipole potential induced by the non-adiabatic part is smaller compared to the non-spinning case but is still responsible for radially outward motion. We take advantage of this ordering to develop a theory where the adiabatic potential is zeroth order and the dipole potential is first order. This theory provides a good explanation of the simulation results and it is able to explain several experimental observations. A transport barrier criteria is obtained from the current theory and the transport barrier condition in the simulation and from experimental data is consistent with the behavior of blobs in the simulation and in experiments. The current theoretical framework can be extended and enriched to include kinetic and other effects, enabling more accurate descriptions of the experimental behavior of blobs in magnetically confined devices.

It is noted that this work does not aim to model the buildup of the whole pedestal. This work aims to isolate the spinning blobs and zonal field, and investigate the zonal field effect on the spinning blobs. Although the assumptions such as the blob seeding and flat profiles are used in our simulations, these gyrokinetic simulations with realistic geometry and equilibrium are first principles with the drift kinetic electrons, including both adiabatic and non-adiabatic electron responses. In the future, we will investigate the turbulent interaction of many  blobs with realistic magnetic geometry and profiles in different parameter regimes. We will also consider different birth rates in L-mode and H-mode, and their impact on the SOL width, leading to a more complete theoretical transport model.  
\\

\textbf{Acknowledgements}
This research was supported by the Exascale Computing Project (17-SC-20-SC), a collaborative effort of the U.S. Department of Energy Office of Science and the National Nuclear Security Administration, and the SciDAC-4 project High-fidelity Boundary Plasma Simulation funded by the U.S. Department of Energy Office of Science under Contract No. DE-AC02-09CH11466. This research used resources from the Oak Ridge Leadership
Computing Facility at the Oak Ridge National Laboratory (ORNL)
and resources of the National Energy Research Scientific Computing Center (NERSC), which is supported by the Office of Science of the U.S. Department of Energy under Contract Nos. DEAC05–00OR22725 and DE-AC02–05CH11231, respectively.

\end{document}